\documentclass{webofc}
\usepackage[varg]{txfonts}   

\begin{document}
\title{The effects of pseudorapidity-dependent observables on (3+1)D Bayesian Inference of relativistic heavy-ion collisions}

\author{\firstname{Chun} \lastname{Shen}\inst{1,2} \fnsep\thanks{Speaker, \email{chunshen@wayne.edu}}
        \and
        \firstname{Bj\"orn} \lastname{Schenke}\inst{3}
        \and
        \firstname{Wenbin} \lastname{Zhao}\inst{1,4}
}

\institute{
    Department of Physics and Astronomy, Wayne State University, Detroit, Michigan 48201, USA
\and
    RIKEN BNL Research Center, Brookhaven National Laboratory, Upton, NY 11973, USA
\and
    Physics Department, Brookhaven National Laboratory, Upton, NY 11973, USA
\and
    Nuclear Science Division, Lawrence Berkeley National Laboratory, Berkeley, California 94720, USA
}

\abstract{%
This proceeding highlights the effects of pseudorapidity-dependent charged hadron observables $dN^\mathrm{ch}/d\eta$ and $v_2^{\rm ch}(\eta)$ in Au+Au collisions at 200 GeV on constraining the initial-state nuclear stopping for the beam remnants and the effective QGP specific shear viscosity in a recent Bayesian inference analysis using an event-by-event (3+1)D hydrodynamics + hadronic transport theoretical framework.
}
\maketitle
\section{Introduction}

Relativistic heavy-ion collisions offer a controlled environment for investigating the emergent properties of deconfined nuclear matter, the Quark-Gluon Plasma (QGP), at extreme temperatures and densities. Recent advancements in detector technologies at the Relativistic Heavy-Ion Collider (RHIC) and the Large Hadron Collider (LHC) have paved the way for more precise measurements, particularly for (pseudo)rapidity-dependent observables. These measurements allow us to systematically study the longitudinal dynamics of relativistic nuclear collisions at various system sizes and as a function of collision energy. Such phenomenological studies push theoretical descriptions into the (3+1)D era~\cite{Achenbach:2023pba,Arslandok:2023utm,Shen:2020mgh}.
In the meantime, the increased complexity of measurements in relativistic heavy-ion collisions and the need for high-precision simulations at a large scale present significant computational challenges. 
One notable approach to extracting meaningful information from the wealth of experimental data is through global multi-dimensional Bayesian inference analyses~\cite{Pratt:2015zsa, Auvinen:2017fjw, Bernhard:2019bmu, Nijs:2020ors, JETSCAPE:2020shq, Parkkila:2021yha, Heffernan:2023gye}. Bayesian methods provide a systematic and statistically rigorous framework for combining experimental results with theoretical models, deriving robust constraints on the physical properties of collision systems. In \cite{Shen:2023awv}, we performed a comprehensive Bayesian analysis, utilizing (3+1)D hydrodynamic simulations, to constrain the initial-state nuclear stopping and the QGP's specific shear and bulk viscosities using measurements from the RHIC Beam Energy Scan (BES) program phase I.
This proceeding provides complementary information on the effect of including the PHOBOS pseudorapidity-dependent measurements in Au+Au collisions at 200 GeV on the posterior constraints obtained from the Bayesian analysis. The comparison highlights that the rapidity-dependent observables can contribute significantly to constraining the (3+1)D collision dynamics and the transport properties of QGP. 

\section{The Model Framework}

We employed the \textsc{3d-Glauber + music + urqmd} hybrid framework~\cite{Shen:2017bsr,Shen:2022oyg, Zhao:2022ayk} to simulate the dynamical evolution of Au + Au collisions event-by-event at RHIC. The  physical properties of the collision systems, such as initial-state nuclear stopping and QGP specific viscosities, are parametrized as model parameters in the theoretical framework. Overall, we designed a 20-dimensional model parameter space for Bayesian inference analysis~\cite{Shen:2023awv}. 
Because this (3+1)D framework is numerically expensive, the designed Bayesian analysis requires us to train fast surrogate emulators for the model results. We generated training data for 1,000 parameter sets and fit them with Gaussian Process (GP) emulators for the experimental observables of interest. Then, we simulated 1,000 minimum bias collision events for every training parameter set to ensure sufficient statistics for those experimental observables. A detailed description of the model can be found in~\cite{Shen:2023awv}.

In this proceeding, we focus on quantifying the impacts of pseudorapidity-dependent measurements of Au+Au collisions at 200 GeV on the Bayesian posterior constraints. We start with a baseline Bayesian posterior distribution using only the mid-rapidity measurements from the STAR Collaborations for Au+Au collisions at 200 GeV, which includes identified particle yields, their averaged transverse momenta, and charged-hadron anisotropic flow coefficients $v_2\{2\}$ and $v_3\{2\}$ from 0 to 60\% in collision centrality~\cite{STAR:2008med, STAR:2017idk, STAR:2016vqt}. Then, we will include step-by-step PHOBOS measurements for the charged hadron pseudorapidity dependent yield $dN^{\rm ch}/d\eta$~\cite{PHOBOS:2005zhy} and elliptic flow $v^{\rm ch}_2(\eta)$~\cite{PHOBOS:2006dbo} in our Bayesian analysis and compare the changes in the resulting posterior distributions. We will focus on the constraints on the initial-state rapidity loss of the beam remnants and the $\mu_B$-dependent QGP specific shear viscosity.

The \textsc{3d-Glauber} model considers that the beam remnants of wounded nucleons lose some amount of energy and momentum, which are then fed to the collision system's hydrodynamic energy-momentum tensor as source currents at forward and backward space-time rapidities. The beam remnants' rapidity loss is parameterized using a model parameter $\alpha_{\rm rem}$, which characterizes the ratio of the amount of energy and momentum loss to that in the string deceleration of the colliding parton pairs~\cite{Shen:2022oyg}. We vary $\alpha_{\rm rem} \in [0, 1]$ in the Bayesian analysis. The limit of $\alpha_{\rm rem} = 1$ means that the beam remnants lose an equal amount of energy and momentum on average compared to those of the colliding partons, while $\alpha_{\rm rem} = 0$ stands for no energy loss. Please refer to Ref.~\cite{Shen:2023awv} for more details.

\section{Results}

\begin{figure}[h!]
    \centering
    \includegraphics[width=0.49\linewidth]{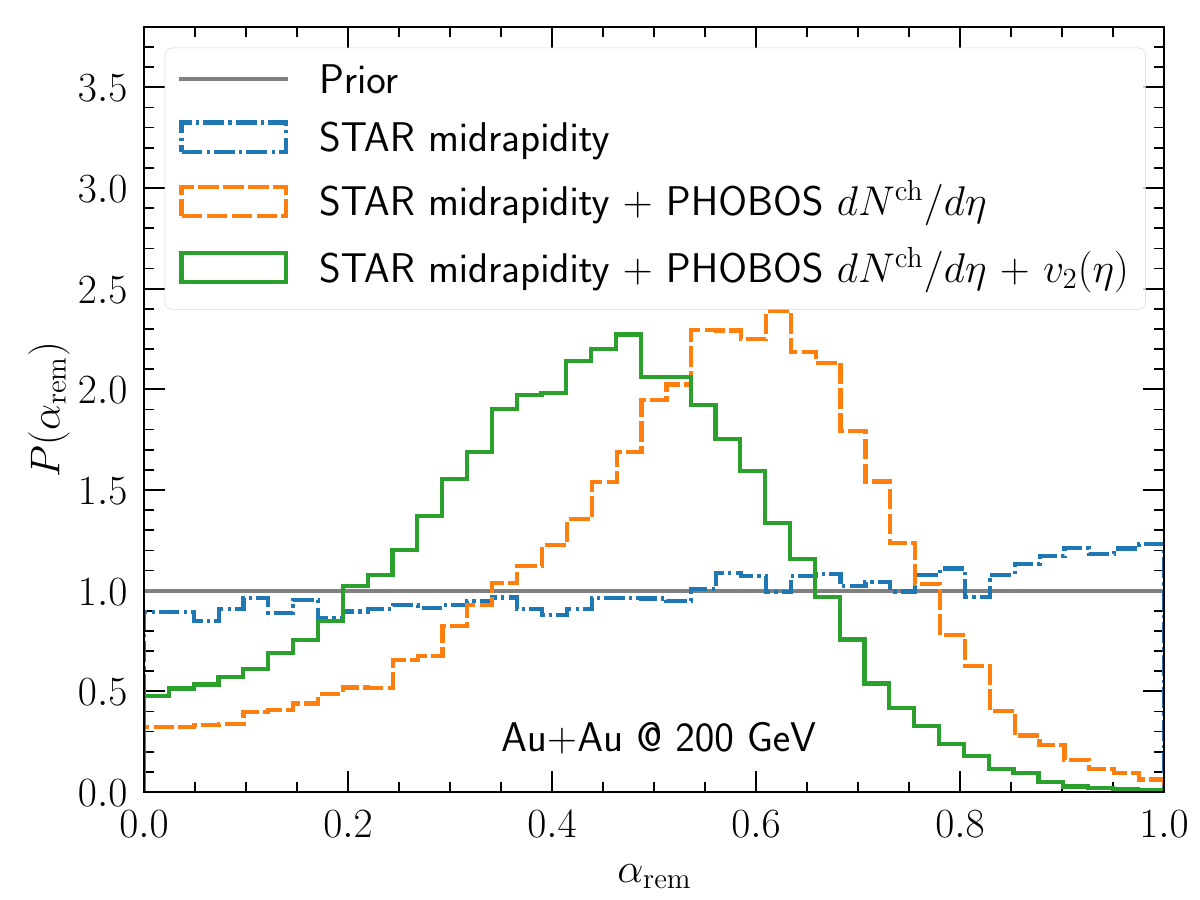}
    \includegraphics[width=0.49\linewidth]{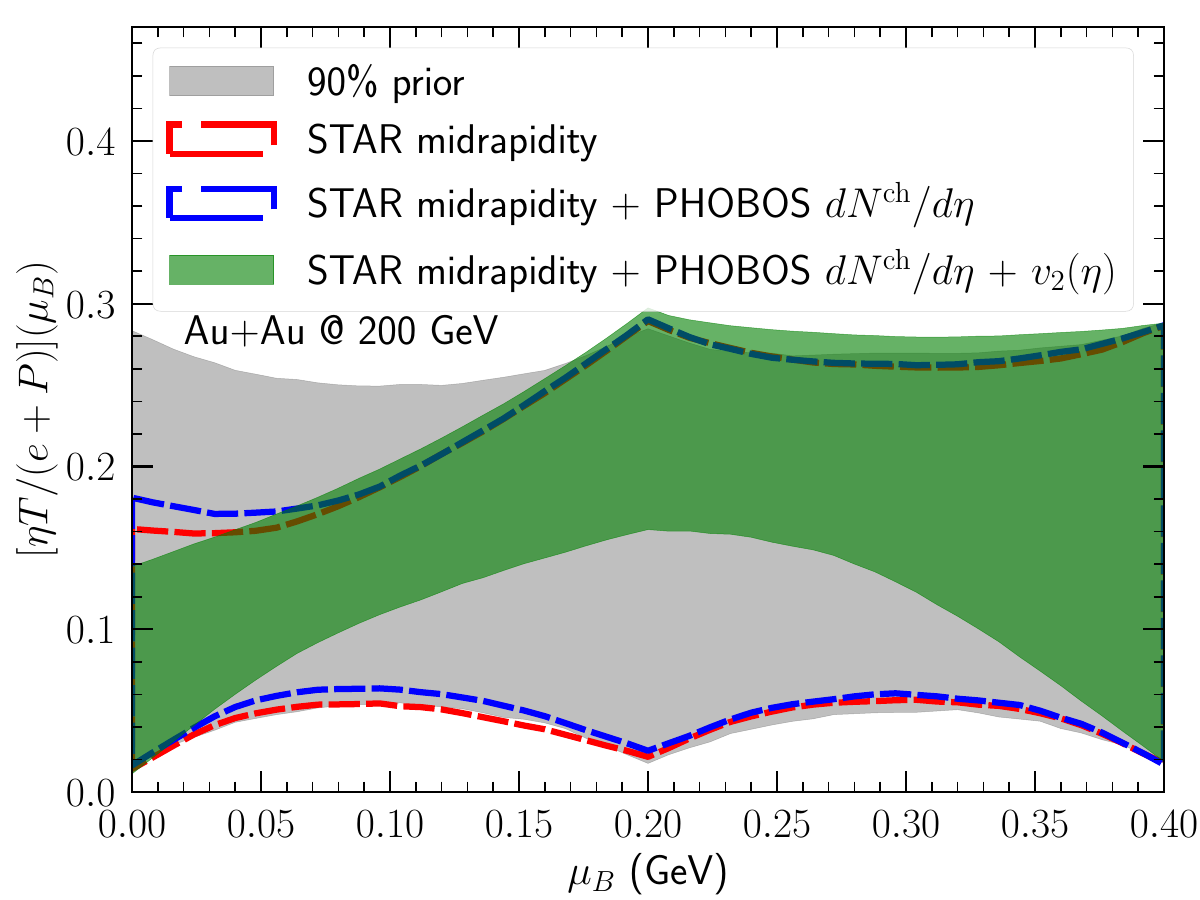}
    \caption{Left Panel: The prior and posterior distributions of the fraction of rapidity loss for the beam remnants. Right Panel: the 90\% prior and posterior bands for the net baryon chemical potential dependence of the QGP effective specific shear viscosity.}
    \label{fig1}
\end{figure}

Figure~\ref{fig1} shows the posterior distributions for the ratio of the beam remnants' rapidity loss to string deceleration and the $\mu_B$-dependent QGP specific shear viscosity by the subsequent inclusion of the PHOBOS pseudo-rapidity measurements.

Because the beam rapidity of the colliding nucleons is $y_{\rm beam} = 5.36$ at $\sqrt{s_{\rm NN}} = 200$\,GeV, midrapidity measurements from the STAR Collaboration are expected to impose little constraint on the amount of energy loss for the beam remnants. This expectation is confirmed by the small difference between the prior and posterior distributions (STAR midrapidity) shown in the left panel of Fig.~\ref{fig1}. Once we include the PHOBOS $dN^{\rm ch}/d\eta$ measurements in the Bayesian analysis, the posterior distribution (the orange dashed line) changes significantly, and $\alpha_{\rm rem} \sim 1$ is strongly disfavored. This result suggests that the average rapidity loss for beam remnants is smaller than that of the colliding partons. The PHOBOS pseudorapidity-dependent elliptic flow $v^{\rm ch}_2(\eta)$ measurements introduce an additional constraint and further shift the posterior distribution towards smaller values of $\alpha_{\rm rem}$. Including constraints from all experimental observables, the most probable value of $\alpha_{\rm rem}$ is around $0.4 \sim 0.5$, which indicates that the beam remnants lose about half of the energy and momentum compared to those of the colliding partons. This Bayesian constraint is consistent with our earlier model study based on measurements of charged hadron $dN^{\rm ch}/d\eta$ in minimum bias p+p collisions~\cite{Shen:2022oyg}.

The right panel of Fig.~\ref{fig1} shows the impacts of PHOBOS measurements on the $\mu_B$-dependent QGP specific shear viscosity. The STAR elliptic flow measurements for Au+Au 200 GeV collisions at midrapidity can constrain the QGP specific shear viscosity around $\mu_B = 0$. These data reduce the 90\% confidence region in the parameter space by approximately half compared to the prior. Adding the PHOBOS $d N^{\rm ch}/d\eta$ measurements does not improve the posterior distribution. This insensitivity is expected because the charged hadron yields are more sensitive to the initial-state rapidity loss parameters in the model than to the QGP shear viscosity.

On the other hand, the PHOBOS $v^{\rm ch}_2(\eta)$ measurements impose significant constraints on the QGP specific shear viscosity at $\mu_B \sim 0.2$\,GeV. The additional constraints at finite $\mu_B$ come from the fact that the fireball medium contains more net baryon density in the forward and backward rapidity regions than at midrapidity. Therefore, the elliptic flow measurements from these rapidity regions can better constrain the QGP specific shear viscosity at finite $\mu_B$. It is worth noting that this constraint complements the midrapidity measurements at lower collision energies, at which more net baryon charged are transported to the midrapidity window. Therefore, we will gain strong constraints on the QGP specific shear viscosity by combining the rapidity-dependent anisotropic flow measurements and their collision energy dependence~\cite{Shen:2018pty,Li:2018fow,Shen:2020jwv}.

\section{Conclusions}

This proceeding provides a focused study on quantifying the impact of including rapidity-dependent measurements in (3+1)D Bayesian analyses to constrain the physics properties of relativistic heavy-ion collisions. Although PHOBOS pseudorapidity measurements contain large relative uncertainties, they provide complementary constraints on beam remnants' rapidity loss and the QGP specific shear viscosity at finite $\mu_B$. By including future precision measurements of rapidity-dependent measurements from the STAR BES phase II program in the Bayesian analysis, we anticipate obtaining robust phenomenological constraints on the QGP's transport properties at finite net baryon density.

\section*{Acknowledgments}
This work is supported by the U.S. Department of Energy (DOE), Office of Science, Office of Nuclear Physics, under DOE Contract No. DE-SC0012704 (B.P.S.) and Award No. DE-SC0021969 (C.S.).
C.S. acknowledges a DOE Office of Science Early Career Award.
W.B.Z. is in part supported by the National Science Foundation (NSF) under grant number ACI-2004571 within the framework of the XSCAPE project of the JETSCAPE Collaboration and DOE Contract No. DE-AC02-05CH11231 within the framework of the Saturated Glue (SURGE) Topical Theory Collaboration. This research was done using computational resources provided by the Open Science Grid (OSG)~\cite{Pordes:2007zzb, Sfiligoi:2009cct}, which is supported by the NSF award \#2030508 and \#1836650.

\bibliography{references}

\end{document}